\documentclass[letterpaper]{article} 
\usepackage{aaai23}  
\usepackage{times}  
\usepackage{helvet}  
\usepackage{courier}  
\usepackage[hyphens]{url}  
\usepackage{graphicx} 
\usepackage{natbib}  
\usepackage{caption} 
\usepackage{appendix}
\usepackage{graphicx}
\usepackage{lineno}
\usepackage{array}
\usepackage{longtable}
\usepackage{multirow}
\usepackage{listings}
\usepackage{color}
\usepackage{graphics}
\usepackage{tabularx}
\usepackage{longtable}
\usepackage{array} 
\usepackage{booktabs}
\usepackage{epstopdf} 
\usepackage{multirow}
\usepackage{pdflscape}
\usepackage{afterpage}
\usepackage{capt-of}
\usepackage{rotating}
\usepackage{algorithmic}
\usepackage{algorithm}
\usepackage[utf8]{inputenc}
\usepackage{placeins}
\usepackage{ragged2e}
\usepackage[export]{adjustbox}
\usepackage{adjustbox}
\usepackage[misc]{ifsym}
\usepackage{notoccite}
\usepackage{xspace}
\usepackage{caption}
\usepackage{subcaption}
\usepackage{amsmath}
\usepackage{amsthm}
\usepackage{cleveref}
\usepackage{xcolor}

\usepackage{newfloat}
\usepackage{listings}
\DeclareCaptionStyle{ruled}{labelfont=normalfont,labelsep=colon,strut=off} 
\floatstyle{ruled}
\newfloat{listing}{tb}{lst}{}
\floatname{listing}{Listing}
\newcommand{\one}{({\em i}\/) \xspace}
\newcommand{\two}{({\em ii}\/) \xspace}

\crefformat{section}{\S#2#1#3} 
\crefformat{subsection}{\S#2#1#3}
\crefformat{subsubsection}{\S#2#1#3}
\crefformat{pb}{\S#2#1#3}

\def\eg{\emph{e.g.} \xspace}
\def\etc{\emph{etc.} \xspace}
\def\ie{\emph{i.e.} \xspace}

\newcommand{\pb}[1]{\vspace{0.75ex}\noindent{\bf \em #1}\hspace*{.3em}}

\ifx\commentsoff\undefined
	\newcommand\waleed[1]{\textbf{\textcolor{green}{WI: #1}}	}
	\newcommand\ignacio[1]{\textbf{\textcolor{blue}{IC: #1}}	}
	
\else
	\newcommand\waleed[1]{}
	\newcommand\ignacio[1]{}
\fi

\def\eg{\emph{e.g.}\xspace}
\def\etc{\emph{etc.}\xspace}
\def\ie{\emph{i.e.}\xspace}

\title{How Similar Are Elected Politicians and Their Constituents? Quantitative Evidence From Online Social Networks}
\author{
    Waleed Iqbal\textsuperscript{\rm 1}, Gareth Tyson\textsuperscript{\rm 1, \rm 2}, Ignacio Castro\textsuperscript{\rm 1}
}
\affiliations{
    \textsuperscript{\rm 1}Queen Mary University of London,\\  \textsuperscript{\rm 2}Hong Kong University of Science and Technology (GZ)\\
\{w.iqbal, i.castro\}@qmul.ac.uk,
gtyson@ust.hk
}

\usepackage{bibentry}
\nocopyright
\begin{document}

\maketitle

\begin{abstract}
How similar are politicians to those who vote for them?
This is a critical question at the heart of democratic representation and particularly relevant at times when political dissatisfaction and populism are on the rise.
To answer this question we compare the online discourse of elected politicians and their constituents.
We collect a 
two and a half years (September 2020 – February 2023)
constituency-level dataset for USA and UK that includes: 
\one the Twitter timelines (5.6 Million tweets) of elected political representatives (595 UK Members of Parliament and 433 USA Representatives),
\two the Nextdoor posts (21.8 Million posts) of the constituency
(98.4\% USA and 91.5\% UK constituencies).
We find that elected politicians tend to be equally similar to their constituents in terms of content and style regardless of whether a constituency elects a right or left-wing politician. 
The size of the electoral victory and the level of income of a constituency  shows a nuanced picture.
The narrower the electoral victory,
the more similar the style and the more dissimilar the content is. 
The lower the income of a constituency,
the more similar the content is. 
In terms of style, 
poorer constituencies tend to have a more similar sentiment and
more dissimilar psychological text traits (\ie measured with LIWC categories).
\end{abstract}

\section{Introduction}
\label{sec:intro}
Dissatisfaction with democracy is reaching an all time high~\cite{foa2020global}
and 
 populism is on the rise~\cite{martelli2023populist, populists-pew}.
Recent survey data shows that voters do not feel well represented by their elected politicians~\cite{views-political-representation}.
\citeauthor{jones2020chavs} \cite{jones2020chavs} argues that left-wing politicians in the UK are increasingly dissimilar to the lower income voters that they claim to represent. Piketty~\cite{piketty2018brahmin} left-wing parties are increasingly failing to represent their traditional lower income  voters  across western countries. 

This disconnect between political representatives and their constituents seems to be at the heart of the growing political distrust. 
Voters worldwide feel that politicians should be more alike to the voters they represent~\cite{representative-views}. 
Evidence from the UK shows that the perception that political representatives have little common ground with voters is indeed a key driver of political distrust~\cite{valgardhsson2021good}.
In this paper we look at this precise issue. 
We study the similarities between political representatives and their constituents by comparing the online discourse of elected representatives and the voters in their respective constituency.
We are particularly interested in analyzing whether the degree of similarity varies depending on factors such as political ideology, income, and the margin of victory in the most recent elections.

To examine this, we collect two large datasets with the online discourse of elected politicians and their constituents, and map the elected politicians to their constituents (\ie who elected them) for most constituencies in USA and UK.
Specifically, we collect \one Twitter timelines of elected politicians, and \two Nextdoor posts from their respective constituents.
Nextdoor is a location-based social network where users interact within closed social networks of neighbors that have validated their home addresses, \ie a constituent posting about a local issue is in fact residing in that constituency.
We leverage this to  map users to their respective constituency.
Our Nextdoor dataset includes 21.8 Million posts from 190,706 neighborhoods (433 (98.4\%) constituencies) in the United States (USA) and 21,046 neighborhoods (595 (91.5\%) constituencies) in the United Kingdom (UK) between September 2020 and February 2023. These constituencies cover 98.6\% population in the USA and 94.6\% population in the UK.

We also collect Twitter timelines of Members of Parliament (MPs) from the UK’s House of Commons and USA Representatives (USAReps) from the USA’s House of Representatives. This includes 5.6 Million tweets from 433 USAReps’ account timelines out of 440 USAReps and 595 UK MPs’ account timelines out of 650 UK MPs.


Using this data, we conduct a constituency-level study:
for each constituency, we compare the online discourse of the elected politician (Twitter) with 
the online discourse of the constituents (Nextdoor) who elected the politician.
We compare the online discourse both in terms of content and style 
by looking at the semantic similarity of the content and psychological traits in the text (LIWC categories)  and text sentiment.
We address the following Research Questions (RQs):

\begin{itemize}
    \item \textbf{RQ1:} Are right-wing politicians more similar to their constituencies than left-wing politicians?
    \item \textbf{RQ2:} Are politicians in more disputed constituencies more similar to their constituents?
    \item \textbf{RQ3:} Are politicians in poorer constituencies more similar to their constituents than in richer ones?
\end{itemize}

Our main findings include:
\begin{itemize}
    \item We find that elected politicians are frequently equally similar in terms of content and style. This is the case regardless of whether they are a right or left-wing politician. 
    The level of similarity is relatively low but higher than when comparing constituents with elected politicians of other constituencies:
    we compare right-wing constituencies with left-wing politicians (and vice versa) and find that 
    the similarity is substantially lower. 
    \item The size of the victory in the elections shows a nuanced picture. 
    We find that narrower electoral victories are associated with a more similar style (both in terms of LIWC categories and sentiment). We find the opposite for content: the larger the victory the more similar the content tends to be.
    \item Income is also related with varying similarities.
    Constituencies with lower income tend to have more similar content.
    In terms of style,
    poorer constituencies tend to have a more similar sentiment and 
    the opposite is true for the psychological markers (\ie LIWC categories).
\end{itemize}

\section{Data and Methodology}
\label{sec:methodology}

\subsection{Datasets}
\label{sec:dataset_nd}

\pb{Nextdoor primer.} Nextdoor is a location-based social network with over 305,000 registered neighborhoods in 11 countries and over 69 million users~\cite{nextdoor-stats}. Nextdoor divides geographical areas into neighborhoods and assigns users to the \textit{neighborhood} where they reside. To ensure that a user is a neighbor of a particular neighborhood, new users validate their home addresses (\eg via regular ``snail'' mail). 

For each neighborhood, Nextdoor creates a dedicated forum where users post and interact (\eg reply, and react to each other's posts).
Users exclusively interact with their neighbors, \ie the users of the neighborhood they are associated with. As a result, the data from a neighborhood exclusively includes the posts of the users who have validated their location in that geographical area.
We refer to the specific areas into which Nextdoor divides a region as \textit{neighborhood}.

\pb{Nextdoor Data Collection.}
We collect 21,845,284 Nextdoor posts from 212,644 neighborhoods between September 2020 to February 2023 using our custom web scrapers. 
Our data includes all neighborhoods of all USA voting states\footnote{We do not include neighborhoods in American Samoa, the U.S. Virgin Islands, Guam, the Northern Mariana Islands, and Puerto Rico.} (17,397,380 posts from 190,761 USA neighborhoods) and UK (4,447,906 posts, 21,883 neighborhoods). 
We also discard a constituency when it does not have a political representative 
(2 vacant USA constituencies with 51 neighborhoods and 3024 Nextdoor posts where representatives passed away) 
or if the representative has no Twitter account 
(55 UK constituencies with 837 neighborhoods and 9098 posts).

%
Our final dataset includes 21,833,162 posts from 211,752 neighborhoods  (Table~\ref{tab:stats_data}). 
Our data has almost 10 times more posts and triples the number of neighborhoods in previous work~\cite{iqbal2023lady} which only included the 10 largest UK cities (15.8\% neighborhoods, 7.96\% posts), 33.7\% USA neighborhoods (12.6\%  posts) in the USA, and 19 months less (November 2020--September 2021).


%

To gather information about Nextdoor neighborhoods and geolocate them, we employ the methodology in~\cite{iqbal2023lady}. 
Similarly, we map each neighborhood to the available lowest geographical granularity in official statistical data: the ZIP code in the USA and the Lower Layer Super Output Area (LSOA) in the UK.

\pb{Constituency Data.}
We map neighborhoods into the official political constituencies according to the data from the USA House of Representatives \cite{hor-usa} and the UK House of Commons~\cite{hoc-uk}. 
We refer to the representative of a constituency as its \emph{elected politician} for both, 
Members (USAReps) from the US's House of Representative\footnote{https://www.house.gov/} and the Members of Parliament (MPs) from UK's House of Commons.\footnote{https://www.parliament.uk/business/commons/}

For each constituency, we combine all the Nextdoor posts of the corresponding neighborhoods and compare them with all the posts of the elected politician for the constituency.



We refer each constituency as left or right-wing based on the political leaning of their elected politician's party.
In our data, there are two USA parties: the right-wing Republican Party and the left-wing Democratic Party.
We find 220 (50.8\%) right-wing constituencies (covering 51.8\% of the USA neighborhoods and 47.3\% of its posts), and  213 (49.2\%) constituencies with 48.2\% USA neighborhoods, and 52.7\% of its posts).
%

In the UK we find 11 political parties with elected politicians.
We classify them into right and left-wing as in~\cite{jolly2022chapel}.
Three small parties are not included in \cite{jolly2022chapel} classification: Social Democratic and Labour Party (2 constituencies, 10,944 posts), Alba Party (2 constituencies, 1,645 posts), and Alliance Party of Northern Ireland (1 constituency, 4,224 posts).
We manually classify these as left-wing by inspecting their corresponding official website \cite{jarrett2016single}.




We find 315 (52.1\%) UK constituencies, 11,224 (53.3\%) neighborhoods and 2,011,624 (45.3\%) posts within right-wing constituencies; and 280 (47.9\%) constituencies, 9,822 (46.7\%) neighborhoods and 2,427,182 (54.7\%) posts within left-wing ones.



\pb{Twitter Data.}
\label{subsec:stats_twitter}
We also collect the Twitter timelines of the elected politicians of each constituency in the dataset using Twitter Academic API.\footnote{https://developer.twitter.com/en/use-cases/do-research/academic-research}
We obtain 1,391,063 tweets from 433 USAReps' account timelines out of 440 USAReps. 
We excluded seven USAReps from our US dataset either represent non-voting estates or have deceased. Therefore, we do not include their constituencies in our dataset. We also collect 4,203,521 tweets from the Twitter timelines of 595 UK MPs. 
We identify 55 MPs without Twitter accounts and, as mentioned before, we do not include these 55 constituencies (837 Nextdoor neighborhoods, 9098 Nextdoor posts) in our dataset. 

Our USA data includes 63.32\% of tweets from left-wing elected politicians and 36.68\% from right-wing ones. 
For the UK, our data contains 66.37\% of tweets by left-wing politicians and 33.63\% from right-wing ones.

\begin{table}
\normalsize
\centering
\begin{adjustbox}{max width=\columnwidth}
\begin{tabular}{|c|c|c|c|}
\hline
\textbf{Attributes}& \textbf{USA}  & \textbf{UK}& \textbf{Total}   \\ \hline
Nextdoor posts                        & 17,394,356       & 4,438,806 & 21,833,162        \\ \hline
Tweets     & 1,391,063  & 4,203,521 & 5,594,584           \\ \hline
Neighborhoods       & 190,706        & 21,046 & 211,752          \\ \hline
Zip codes (USA)/LSOAs (UK)      & 38,497 & 16,235          &54,732  \\ \hline
Neighbors     & 48,602,160 & 9,744,948 & 58,347,108            \\ \hline
    Constituencies      & 433 & 595 & 1028  \\ \hline
    Twitter accounts (elected  politicians)     & 433 & 595 & 1028            \\ \hline

\end{tabular}
\end{adjustbox}
\caption{Nextdoor and Twitter Dataset (after data cleaning).}
\label{tab:stats_data}
\end{table}

\subsection{Data Augmentation}
\label{sec:data_aug}

\pb{Income and population.}
For each neighborhood, we collect socioeconomic data at the constituency level from the official statistics. 
For the USA constituencies, we obtain the population and median annual income from the latest census~\cite{census-usa}. 
For the UK, we obtain the population and median annual income from the UK's Office of National Statistics from the latest Census update~\cite{population-uk}. 

\pb{Political and polling data.}
We collect the constituency name, Twitter username, party affiliation, and polling results from the House of Representatives~\cite{hor-usa} and the House of Commons~\cite{hoc-uk}.
Elected politicians and constituents might have a different online discourse depending on how disputed in the polls their constituency is. To assess this, 
we rank constituencies based on the size of the electoral victory of the winning candidate over its immediate competitor and calculate deciles of the size of the victory. 
The first decile corresponds to the least disputed constituencies (\ie landslide victory) and the tenth one to the most disputed ones (\ie narrow victory).

\pb{Text embeddings.}
To investigate whether the text posted differs across constituencies, we obtain semantic features of the posts via embedding. Prior to obtaining vector embeddings of our text, we pre-process our Nextdoor and Twitter datasets (\eg removing mentions, URLs, \etc).
For each constituency, we combine all the posts of all the respective neighborhoods.
We then convert each post's text into a single vector embedding using the pre-trained sentence transformer model \textit{all-mpnet-base-v2}.
This model is tuned to map every sentence or short paragraph (up to 384 tokens) to a 768-dimensional vector space while preserving relevant text features \cite{song2020mpnet}. 

\begin{figure}[H]
	\begin{center}
		\includegraphics[width=0.9\columnwidth]{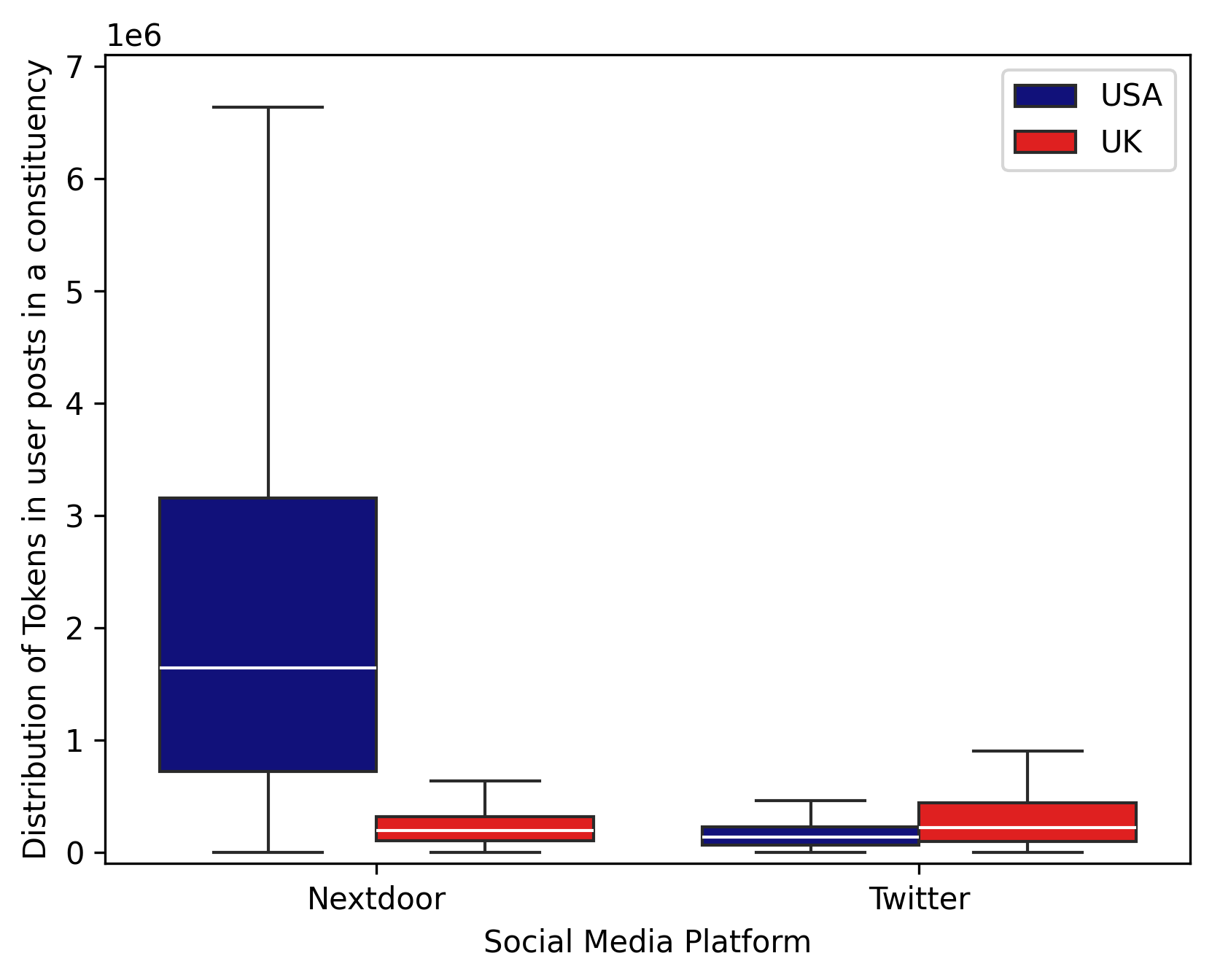}
	\end{center}
	\caption{Distribution of tokens from social media text in USA and UK constituencies in Twitter and Nextdoor after pre-processing.}
	\label{fig:token-dist}
\end{figure}

The number of tokens for all posts in a constituency is high and we cannot directly employ \textit{all-mpnet-base-v2}. 
The number of tokens for the constituency are in higher orders of magnitude. 
The median number of tokens across constituencies in the Twitter data is 132,153 and 216,956 for the USA and UK, 
and for the Nextdoor data they are even larger with a median of 1,640,203 and 191,990 tokens in USA and UK. These numbers even exceed the maximum sequence length  size for highest maximum sequence length sentence embedding model, which is 8096 tokens per input \cite{gunther2023jina} and of our model (maximum sequence length of up to 384 tokens per input). 
Figure \ref{fig:token-dist} shows the distribution of pre-processed tokens in each constituency, dataset, and country. 

To deal with this challenge, 
we compute text embedding of each post in constituency and then aggregate the textual embedding of all posts of a constituency with pair-wise mean-pooling aggregation. 
Mean-pooled textual embedding can represent the complete text corpus into single embedding while maintaining the context of text~\cite{Arora2017,oh2023tadse, singh2022niksss}.
A benefit of this approach, is the weighted importance of content discussed more frequently. 
We finally obtain a 768 dimensional (mean-pooled) embedding for the Nextdoor and Twitter data in each constituency.

\pb{LIWC Categories.}
Linguistic Inquiry and Word Count (LIWC) \cite{pennebaker2015development} is a lexicon-based tool that measures psychologically relevant dimensions in text. 
These dimensions capture both linguistic  (\eg personal pronoun usage) and psychological aspects (\eg effective and social processes) which have been externally validated (\eg measurement of emotions \cite{kahn2007measuring} and social hierarchies \cite{kacewicz2014pronoun}). The LIWC categories consist of 117 categories including 11 root categories, 8 summary variables categories, and 98 subcategories within the root ones~\cite{boyd2022development}. 
We merge all posts from a constituency into a single corpus of text and obtain LIWC scores on the constituency level using 
the LIWC 2022 dictionary (academic license).\footnote{https://www.liwc.app/}
Note that for ease of comparison, we re-scale the LIWC scores from 0--99 to 0-1 so they are comparable with the rest of our results.

\pb{Post sentiment.} 
We label each post's sentiment with a pre-trained Valence Aware Dictionary and Sentiment Reasoner (VADER) model~\cite{hutto2014vader}.
VADER outperforms the typical human reader for social media data (VADER's F1=0.96, Human F1=0.84) \cite{hutto2014vader}.
This also allows for comparison with earlier results with Nextdoor data used this same model~\cite{iqbal2023lady}.

\begin{figure}[H]
	\begin{center}
		\includegraphics[width=0.9\columnwidth]{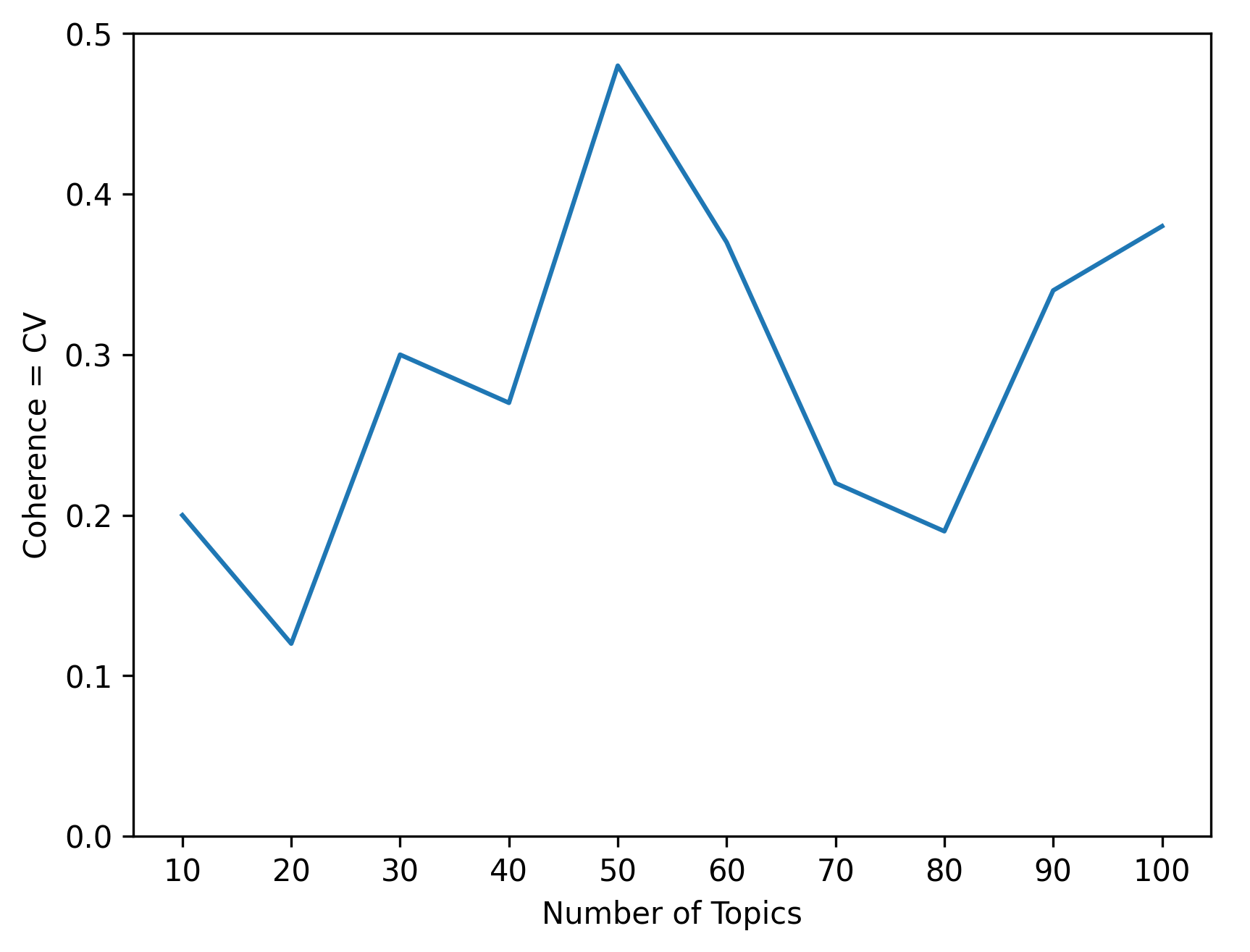}
	\end{center}
	\caption{Coherence Score for BERTopic on Nextdoor and Twitter Datasets.}
	\label{fig:coherence-score}
\end{figure}

\pb{Topic Modeling.}
We identify topics in our data using BERTopic~\cite{huggingface}. 
BERTopic relies on pre-trained transformer-based language models to build document embeddings. It then clusters these embeddings and generates topic representations using the class-based TF-IDF technique~\cite{grootendorst2022bertopic}. 
We use HDBSCAN clustering \cite{campello2013density} for our BERTopic model and combine the Twitter and Nextdoor data. We calculate our BERTopic model on the different number of topics (1--100) achieving the highest coherence score (0.48) on 50 topics (Figure \ref{fig:coherence-score}).

\begin{figure*}[!h]
	\centering
	\begin{subfigure}[b]{\columnwidth}
		\centering
		\includegraphics[width=\linewidth]{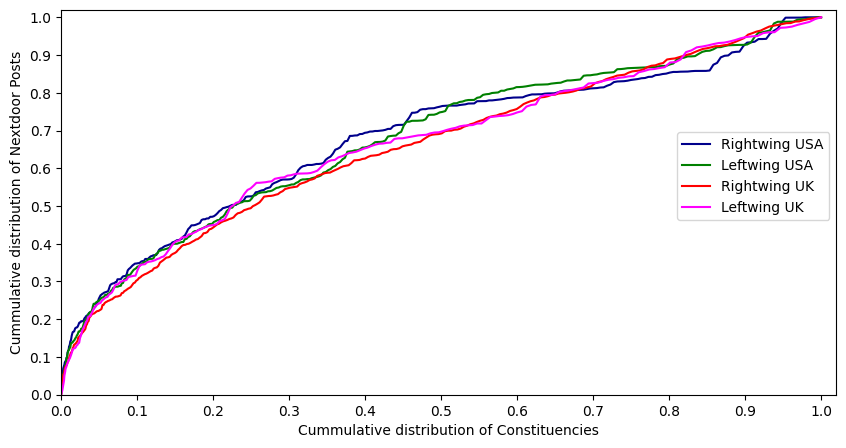}
		\caption{Nextdoor}
		\label{fig:usa_all_embeddings}
	\end{subfigure}
	\hfill
	\begin{subfigure}[b]{\columnwidth}
		\centering
		\includegraphics[width=\linewidth]{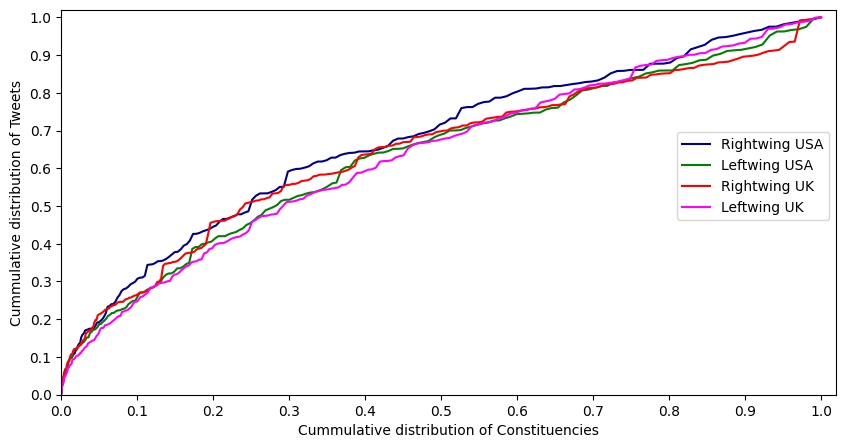}
		\caption{Twitter}
		\label{fig:uk_all_embeddings}
	\end{subfigure}
	\hfill
		\caption{Cumulative distribution of Nextdoor and Twitter data across constituencies in USA and UK.}
\label{fig:all_embeddings}
\end{figure*}

\subsection{Population and data coverage}
Our data includes 98\% constituencies in USA and 92\% constituencies in UK (and all zipcode/LSOA for each constituency). 
This provides comprehensive geographical coverage and substantially extends the data in~\cite{iqbal2023lady}.
Figure~\ref{fig:all_embeddings} shows the cumulative distribution of posts (Nextdoor) and tweets (Twitter) at the constituency level for USA and UK.
We observe similar distributions across both countries with 40\% of the constituencies contribute around 59\%  and 61\% of the Nextdoor posts in USA and UK, respectively. 

We observe similar trends for tweets with 40\% 
elected politicians  responsible for 
about 61\% and 66\% of the tweets in USA and UK, respectively.
\begin{table}[!pbt]
\normalsize
\centering
\begin{adjustbox}{max width=\linewidth}

\begin{tabular}{|l|l|l|l|l|l|l|l|l|} 
\hline
                        & \multicolumn{2}{c|}{\textbf{Population}} & \multicolumn{2}{c|}{\textbf{Nextdoor Posts}} & \multicolumn{2}{c|}{\textbf{Tweets}}& \multicolumn{2}{c|}{\textbf{Nextdoor Users}}  \\ 
\hline
                        & USA  & UK                                & USA  & UK                                    & USA  & UK& USA  & UK                             \\ 
\hline
\textbf{Population}     & 1    & 1                                 & 0.87 & 0.92                                  & 0.92 & 0.88 & 0.83 & 0.81                           \\ 
\hline
\textbf{Nextdoor Posts} & 0.87 & 0.92                              & 1    & 1                                     & 0.90 & 0.89 & 0.93 & 0.88                           \\ 
\hline
\textbf{Tweets}         & 0.92 & 0.88                              & 0.90 & 0.89                                  & 1    & 1 & N/A    & N/A                              \\
\hline
\textbf{Nextdoor Users}         & 0.83 & 0.81                              & 0.93 & 0.88                                  & N/A    & N/A & 1    & 1                             \\

\hline
\end{tabular}
\end{adjustbox}
\caption{Correlation between posts, population, neighborhoods, neighbors, and official population.}
\label{tab:corr_usa_uk_posts}
\end{table}
We further investigate how well our data covers the underlying population by calculating the Pearson correlation between the population of constituencies, tweets and the Nextdoor posts and users in Table~\ref{tab:corr_usa_uk_posts}. 
Similarly to~\cite{iqbal2023lady}, we observe  high  correlation coefficients (above 0.8), giving us confidence in the ability of our data to reflect the underlying population.

\section{Content Similarity}
We first wonder up to what extent elected politicians and constituents discuss the same things.
We first compare the topics they both discuss  and then compare how similar their discourse is by comparing the embeddings of their respective posts.

\subsection{Topics similarity}
\label{sec:topic-modeling}

\begin{figure*}[!h]
	\centering
	\begin{subfigure}[b]{0.9\columnwidth}
		\centering
		\includegraphics[width=\linewidth]{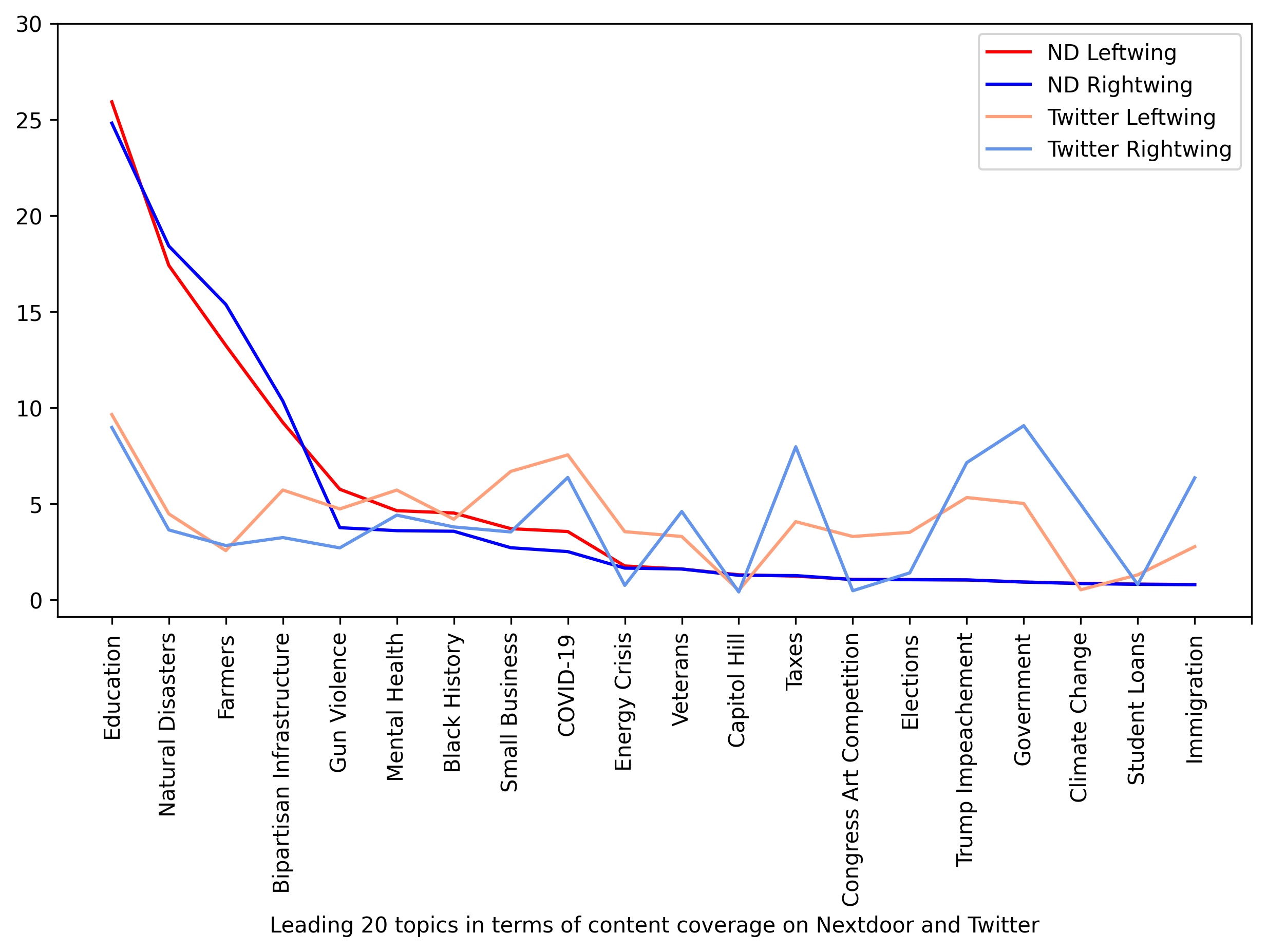}
		\caption{USA}
		\label{fig:usa_topics}
	\end{subfigure}
	\hfill
	\begin{subfigure}[b]{\columnwidth}
		\centering
		\includegraphics[width=\linewidth]{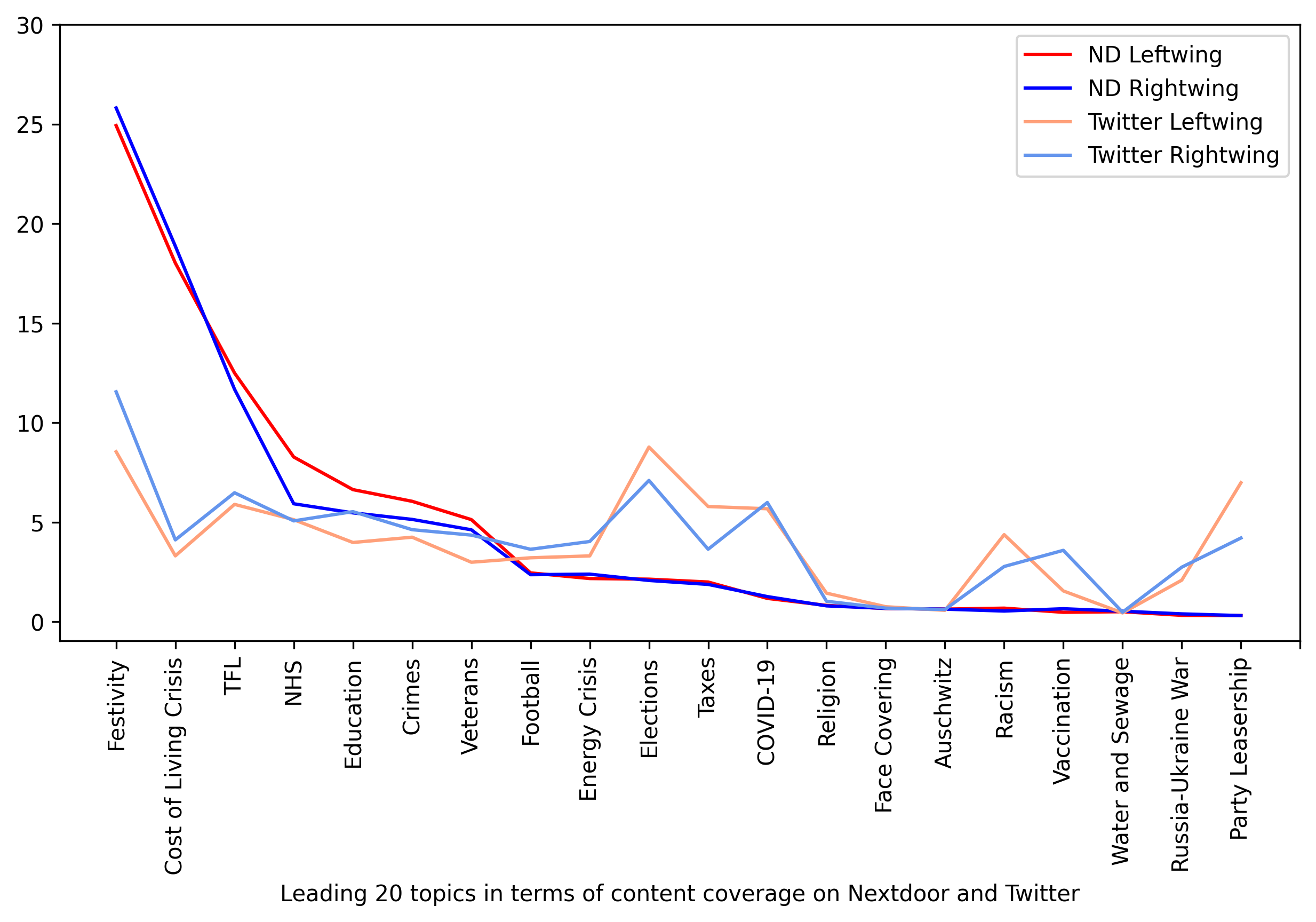}
		\caption{UK}
		\label{fig:uk_topics}
	\end{subfigure}
	\hfill
	\caption{Distribution of posts/tweets by topics over Nextdoor and Twitter.}
	\label{fig:topics_distribution}
\end{figure*}

We first examine which topics are discussed by elected politicians and constituents across different constituencies. 
Figure~\ref{fig:topics_distribution} presents the 20 most discussed topics for USA and UK. 
These top 20 topics cover more than 85\% content discussed over Nextdoor and Twitter (\ie the remaining 30 topics cover just 15\% of the content).

\pb{Elected politicians and their constituents discuss similar topics across the political spectrum.}
The most discussed topics are similar regardless of the political color of a constituency.
This is true for both elected politicians (in Twitter) and constituents (in Nextdoor), although in a slightly different order and size. 
``Education'' and ``Natural Disasters'' are the most discussed topics in online discourse of elected politician (25.3\% and 18.2\% of total tweets) and constituents (9.8\% and 4.3\% of total ND posts) in USA. We observe similar case in UK where ``Festivity'' and ``Cost of Living Crisis'' are most discussed topics by elected politicians (25.1\% and 20.4\% of total tweets) and constituents (10.6\% and 4.9\% of total ND posts).
The topics differ across both countries albeit with some overlaps (\ie education, Covid-19, Energy-crisis, Veterans, Taxes, Elections).

To investigate this further, for each constituency we compare elected politicians and their constituents by  
calculating the cosine similarity of the topics discussed, weighted by their occurrence.
Table~\ref{tab:topics_stats} shows the mean cosine similarity between topic embeddings across constituencies in USA and UK.
We find that elected politicians and constituents discuss similar topics in their constituency regardless of whether a constituency is right or left-wing. 
The close-to-zero variance indicates that this is the case for most constituencies.



\pb{Elected politicians and their constituents discuss similar topics regardless of how contested a constituency is or the level of income.}
Elected politicians and constituents discuss similar topics regardless of the voting majority and the income level of a constituency.
Table \ref{tab:idr_cos} shows how the similarity of topics used by elected politicians and constituents varies depending of the income of a constituency, and
the winning margin of the elected politician in the last elections.
Using results from table \ref{tab:idr_cos}, we show this variation with the Inter-Decile Range (IDR) that compares the
the constituencies with the highest (9th decile) and lowest (1st decile) income/winning majority. -ve/+ve sign with value shows decreasing/increase trend towards 9th decile.

We observe that cosine similarities of topics discussed in right and left-wing constituencies are very close. Results from Table \ref{tab:topics_stats} shows this trend is similar in both the USA (IDR based on Winning Majority and income for right-wing: -0.0058; 0.002, IDR based on Winning Majority and income for left-wing: -0.0049; 0.003) and the UK (IDR based on Winning Majority and income for right-wing: 0.011; 0.004, IDR based on Winning Majority and income for left-wing: 0.016; -0.007).

\begin{table}[!h]
\begin{adjustbox}{max width=\linewidth}
\begin{tabular}{|c|cc|cc|}
\hline
\multicolumn{1}{|l|}{}  & \multicolumn{2}{c|}{\textbf{USA}}                           & \multicolumn{2}{c|}{\textbf{UK}}                            \\ \hline
                        & \multicolumn{1}{c|}{\textbf{Left-wing}} & \textbf{Right-wing} & \multicolumn{1}{c|}{\textbf{Left-wing}} & \textbf{Right-wing} \\ \hline
\textbf{Mean}           & \multicolumn{1}{c|}{0.5848}           & 0.5920          & \multicolumn{1}{c|}{0.4963}          & 0.4952           \\ \hline
\textbf{Std. Deviation} & \multicolumn{1}{c|}{0.0148}          & 0.0098           & \multicolumn{1}{c|}{0.0043}          & 0.0071           \\ \hline
\textbf{Variance}       & \multicolumn{1}{c|}{0.0002}          & 0.0001       & \multicolumn{1}{c|}{0.0001}         & 0.0001          \\ \hline
\end{tabular}
\end{adjustbox}
\caption{Descriptive statistics for cosine similarity of topics discussed by constituents and elected politicians in different constituencies.}
\label{tab:topics_stats}
\end{table}



\subsection{Semantic similarity}
\label{subsec:semantic}
%
%
We were expecting to observe clear differences in the topics discussed by elected politicians and their constituents depending on ideology, income or the size of the winning majority.
To our surprise, the  data revealed the opposite.

\begin{figure}[!h]
	\begin{center}
		\includegraphics[width=0.9\columnwidth]{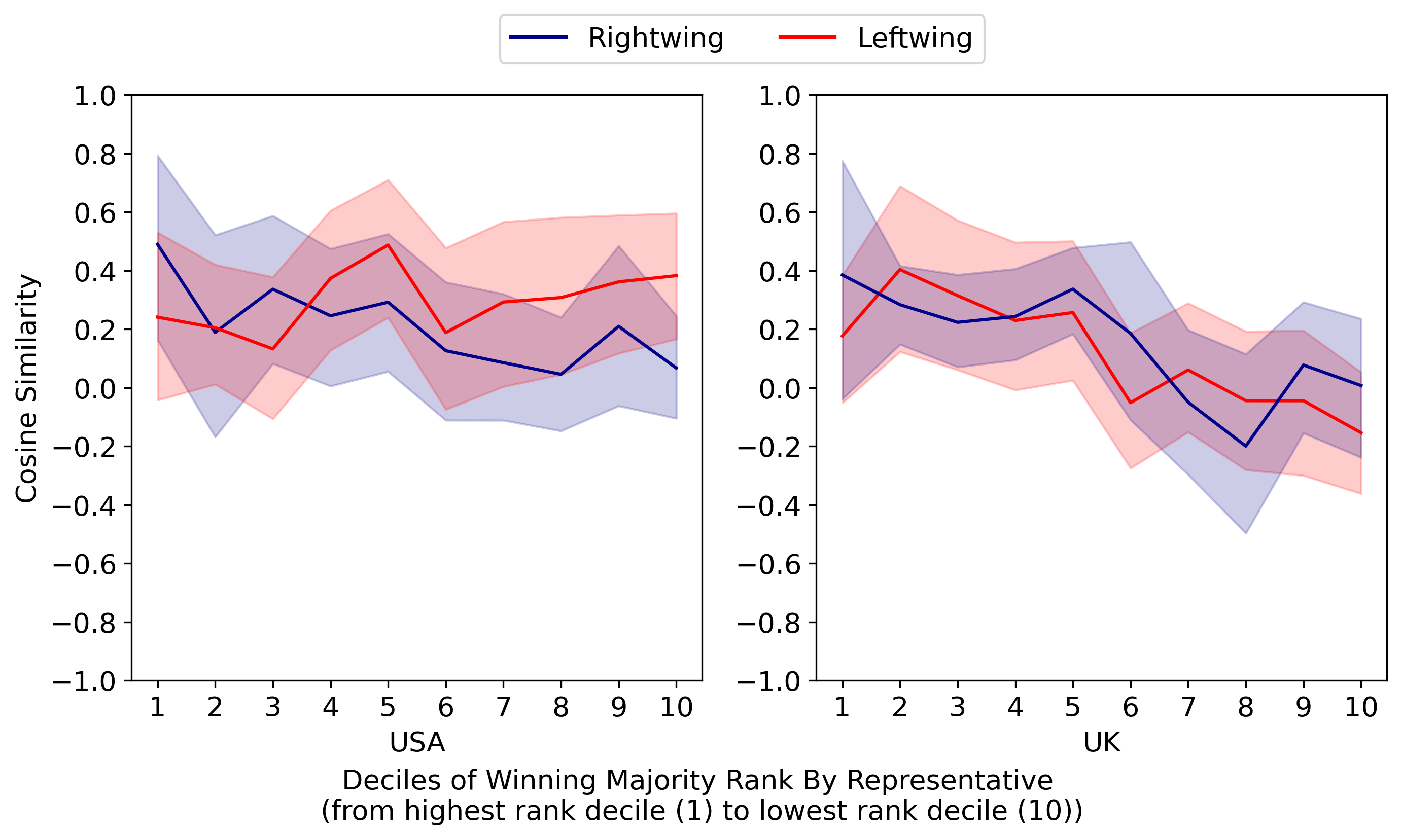}
	\end{center}
	\caption{Distribution of cosine similarity of mean-pooled textual embeddings between constituents and elected politicians over deciles of winning vote majority in different constituencies (from higher to lower).}
	\label{fig:joint_pol_line}
\end{figure}

\begin{figure}[!h]
	\begin{center}
		\includegraphics[width=\columnwidth]{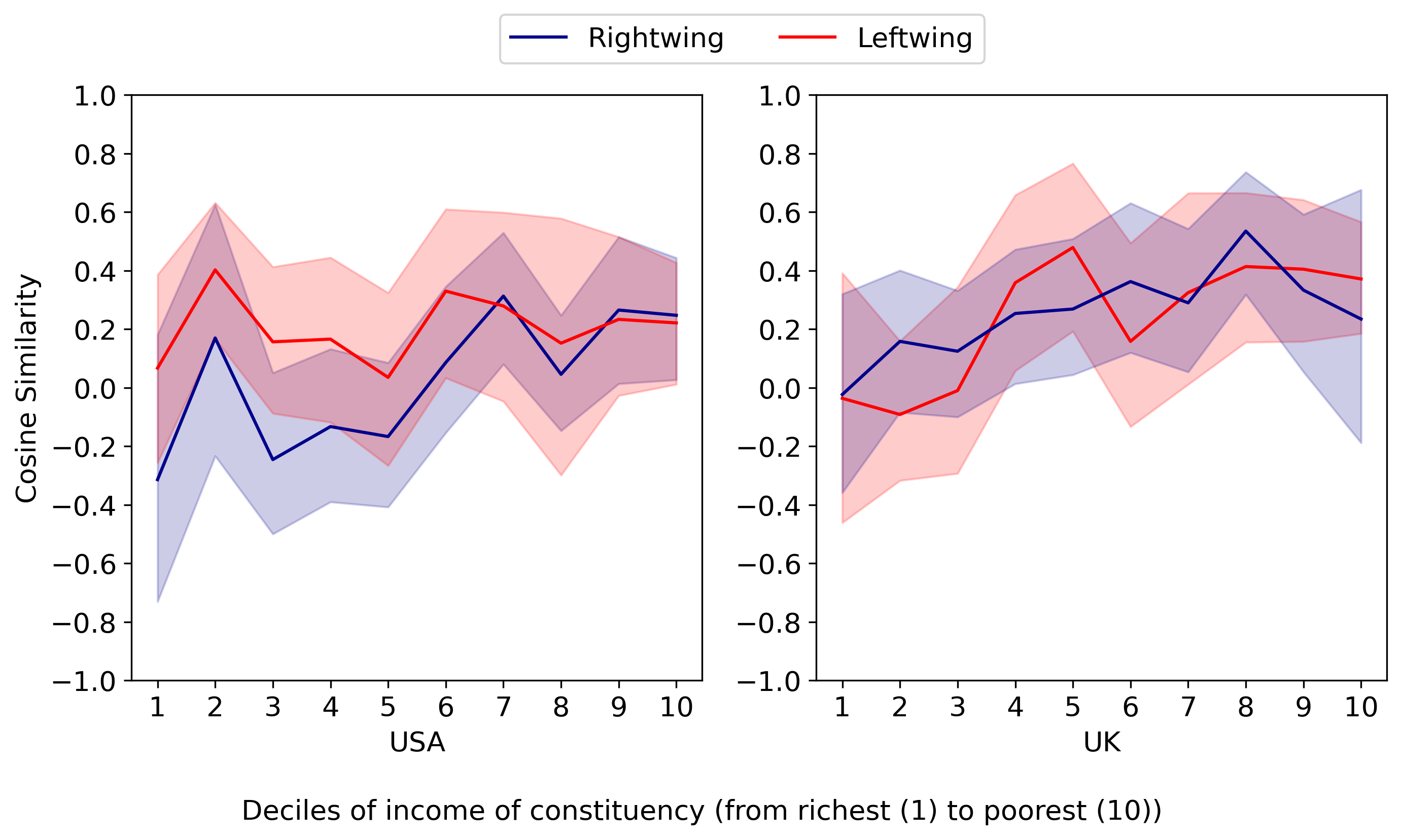}
	\end{center}
	\caption{Distribution of cosine similarity of mean-pooled textual embeddings between Nextdoor and Twitter data over deciles by income levels in different constituencies (from richest to poorest).}
	\label{fig:text_income_line_diff}
\end{figure}

We investigate the issue further by directly measuring the similarity of the content posted by elected politicians and constituents by calculating the cosine similarity between the embeddings of their respective posts on Twitter and Nextdoor. Figure~\ref{fig:joint_pol_line} and Figure~\ref{fig:text_income_line_diff} show the median cosine similarity (line) with overall range of cosine similarity scores (shaded region around line) across USA and UK constituencies.
We plot the cosine similarity for constituencies depending on the size of the electoral victory in Figure~\ref{fig:joint_pol_line} and for different levels of income in Figure~\ref{fig:text_income_line_diff}.
The highest decile (10) corresponds to the most disputed constituencies in 
Figure~\ref{fig:joint_pol_line} and to the poorest ones in Figure~\ref{fig:text_income_line_diff}. 

\begin{table*}[!htbp]
\begin{adjustbox}{max width=\linewidth}
\begin{tabular}{|c|cccc|cccc|}
\hline
\multirow{4}{*}{}  & \multicolumn{4}{c|}{\textbf{Winning Majority}}                                                                                                 & \multicolumn{4}{c|}{\textbf{Income}}                                                                                                           \\ \cline{2-9} 
                   & \multicolumn{2}{c|}{\textbf{USA}}                                                & \multicolumn{2}{c|}{\textbf{UK}}                            & \multicolumn{2}{c|}{\textbf{USA}}                                                & \multicolumn{2}{c|}{\textbf{UK}}                            \\ \cline{2-9} 
                   & \multicolumn{1}{c|}{\textbf{Left-wing}} & \multicolumn{1}{c|}{\textbf{Right-wing}} & \multicolumn{1}{c|}{\textbf{Left-wing}} & \textbf{Right-wing} & \multicolumn{1}{c|}{\textbf{Left-wing}} & \multicolumn{1}{c|}{\textbf{Right-wing}} & \multicolumn{1}{c|}{\textbf{Left-wing}} & \textbf{Right-wing} \\ \cline{2-9} 
                   & \multicolumn{1}{c|}{\textbf{IDR}}      & \multicolumn{1}{c|}{\textbf{IDR}}       & \multicolumn{1}{c|}{\textbf{IDR}}      & \textbf{IDR}       & \multicolumn{1}{c|}{\textbf{IDR}}      & \multicolumn{1}{c|}{\textbf{IDR}}       & \multicolumn{1}{c|}{\textbf{IDR}}      & \textbf{IDR}       \\ \hline
\textbf{Topics}     & \multicolumn{1}{c|}{-0.0049}           & \multicolumn{1}{c|}{-0.0058}            & \multicolumn{1}{c|}{0.016}             & 0.011              & \multicolumn{1}{c|}{0.003}             & \multicolumn{1}{c|}{0.002}              & \multicolumn{1}{c|}{-0.007}            & 0.004              \\ \hline
\textbf{Posts}   & \multicolumn{1}{c|}{0.27}              & \multicolumn{1}{c|}{-0.23}              & \multicolumn{1}{c|}{-0.26}              & -0.34              & \multicolumn{1}{c|}{0.42}              & \multicolumn{1}{c|}{0.58}               & \multicolumn{1}{c|}{0.67}              & 0.41               \\ \hline
\textbf{Sentiment} & \multicolumn{1}{c|}{-0.042}            & \multicolumn{1}{c|}{-0.057}             & \multicolumn{1}{c|}{-0.041}            & -0.039             & \multicolumn{1}{c|}{-0.022}            & \multicolumn{1}{c|}{-0.016}             & \multicolumn{1}{c|}{-0.061}            & -0.067             \\ \hline

\end{tabular}
\end{adjustbox}
\caption{Inter-decile range for cosine similarity for topics and posts, and absolute differences of compound sentiment scores between constituents and elected politicians across different deciles of winning majority and income.}
\label{tab:idr_cos}
\end{table*}

\pb{Right and left-wing elected politicians discuss similar issues to their constituents.}
%
The similarity of what elected politicians and constituents discuss changes little across the political spectrum.
We observe that similarity is limited 
with an overall average cosine similarity score of $0.33$ in the USA and $0.27$ in the UK. 
However, this similarity varies scarcely across the ideological spectrum:
the average similarity between right and left-wing constituencies is $0.28$ and $0.22$ in USA and $0.29$ and $0.23$ UK. The differences between cosine similarity of overall data and data from right-wing and left-wing constituencies are 0.05 and 0.11 in the USA and  0.02 and 0.04 in the UK. 


\pb{Elected politicians in less disputed constituencies have a more similar discourse to their constituents.}
Figure~\ref{fig:joint_pol_line} shows the distribution of similarities of mean-pooled textual embeddings, \ie computed for each constituency by comparing constituents (Nextdoor posts) and elected politicians (Twitter).
We plot this from higher to lower deciles of the winning majority, \ie constituencies in the 10th decile have the narrowest victory and those in the 1st decile represent the largest victory.  
We find that less disputed constituencies tend to have more similarity between constituents and elected politicians.
This is particularly in the UK and right-wing USA constituencies with an average Interdecile Range (IDR) of $-0.27$ and 
exactly the opposite ($0.27$) for USA left-wing constituencies.

A possible reason for this trend is that politicians that succeed in being elected, have a more similar style to their constituents. However, without data from the competing candidates it is not possible to validate the hypothesis.

\pb{Elected politicians of poorer constituencies have more similar discourse to their constituents.}
We observe in Figure~\ref{fig:text_income_line_diff} that both in USA and UK, both in right and left-wing constituencies, there is a clear trend towards greater similarity as the level of income of a constituency decreases (average IDR=$0.52$).
Interestingly, left-wing and right-wing constituencies have analogous similarities (average of $0.28$ and $0.22$ in USA and $0.29$ and $0.23$ UK).
%


\pb{Elected politicians have higher similarity with constituents that they represent than with those that they do not.}
So far we have compared elected politicians with their respective constituents. 
We now investigate whether politicians from right-wing constituencies are similar in discourse to constituents from left-wing constituencies and vice versa.
We compute the cosine similarity of the online discourse between elected politicians and constituents from opposing political sides, based on winning majority and income. 
For this analysis and differently from the previous, there is no pair-wise mapping between elected politicians and constituents. 
Instead, we aggregate our data based on different deciles of winning majority and income. 
We then compute the cosine similarity between the posts of the elected politicians and constituents from different constituencies in each decile.

\begin{figure}[H]
	\begin{center}
		\includegraphics[width=0.9\columnwidth]{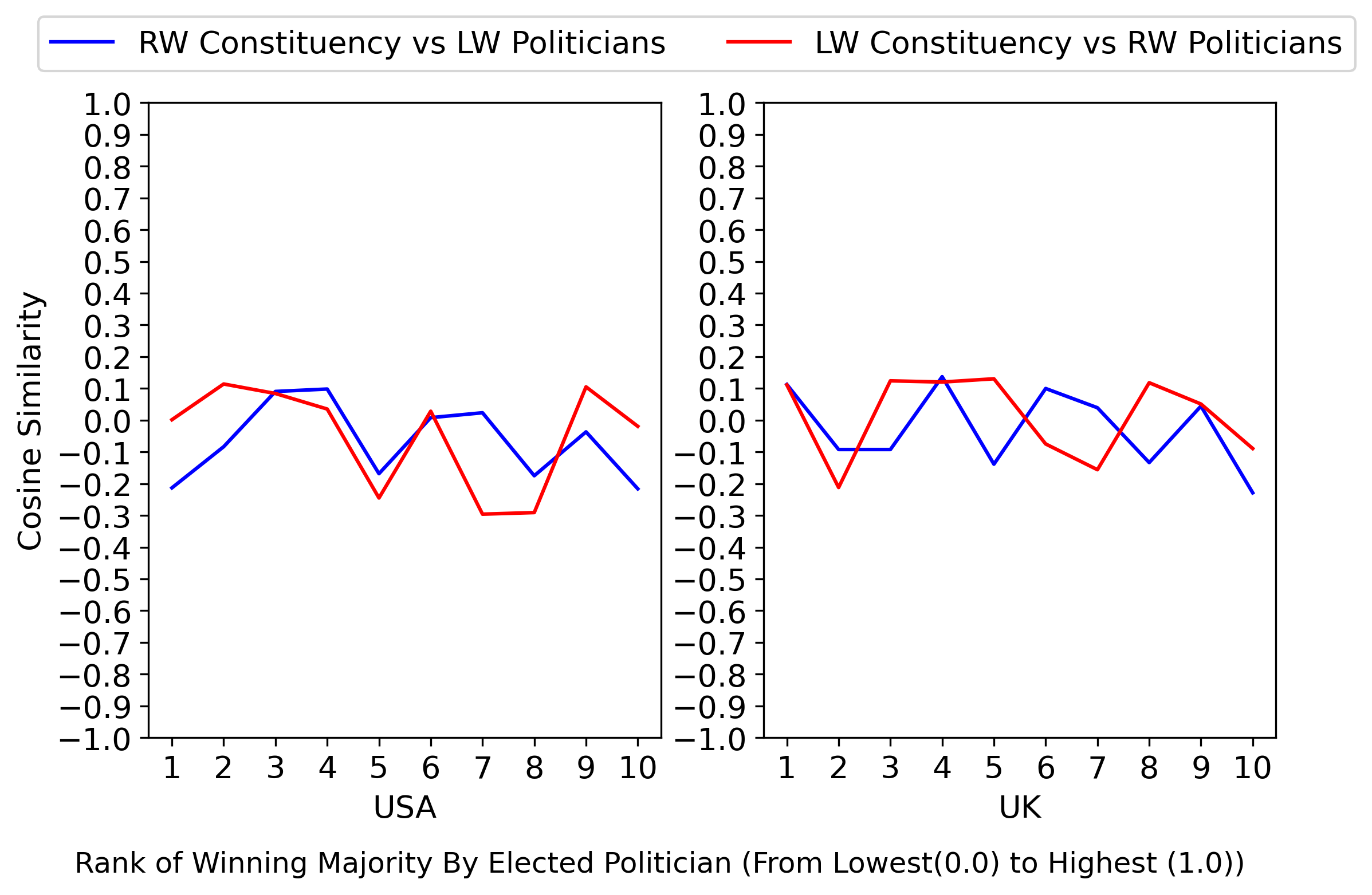}
	\end{center}
	\caption{Distribution of cosine similarity of mean-pooled
textual embeddings between left-wing constituents and right-wing elected politicians and vice versa over deciles of winning vote majority in different constituencies (from higher to lower).}
	\label{fig:comp-winning}
\end{figure}

\begin{figure}[H]
	\begin{center}
		\includegraphics[width=0.9\columnwidth]{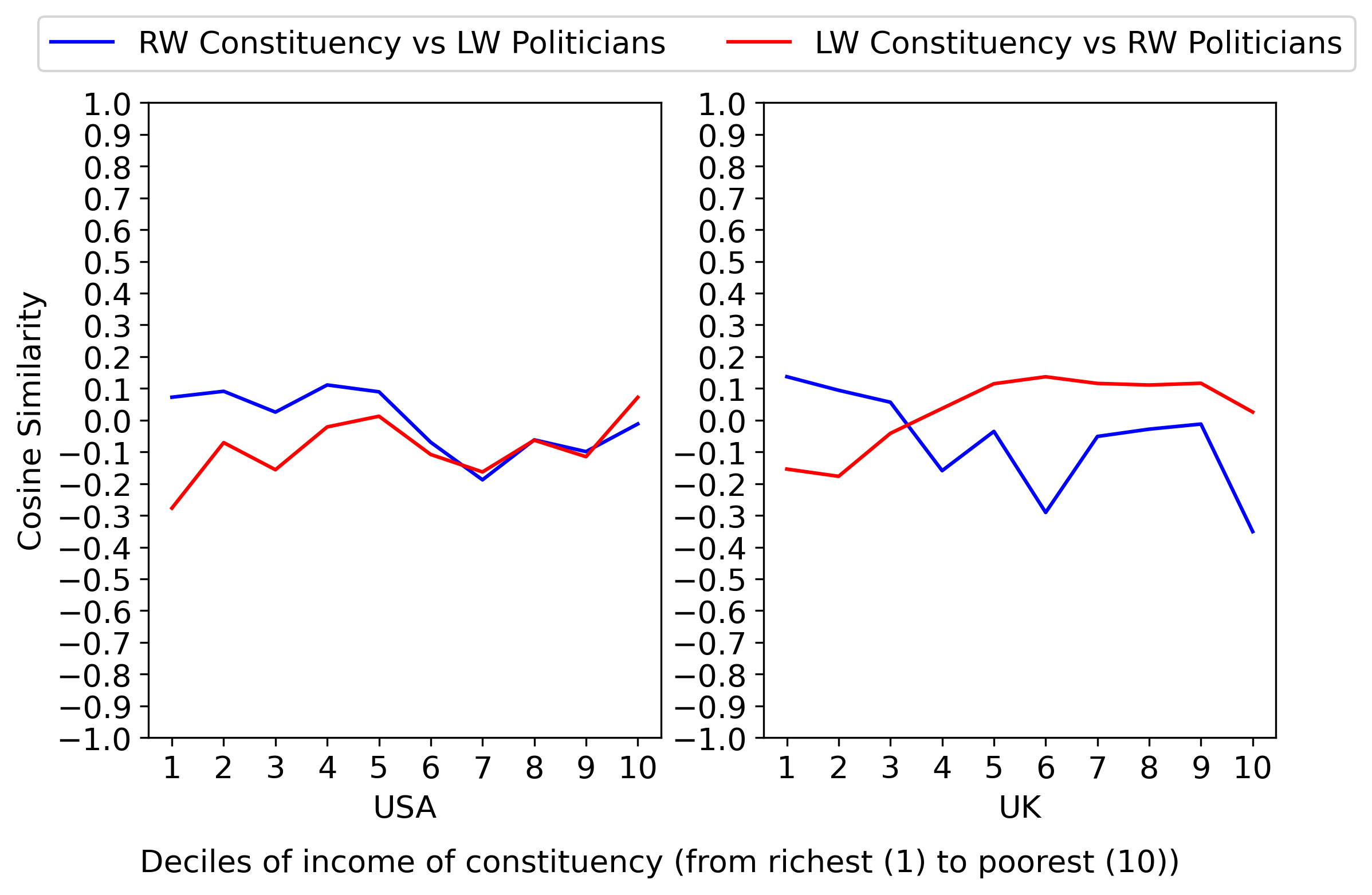}
	\end{center}
	\caption{Distribution of cosine similarity of mean-pooled
textual embeddings between left-wing constituents and right-wing elected politician and vice versa over deciles by income levels in different constituencies (from richest to poorest).}
	\label{fig:comp-income}
\end{figure}

Figures \ref{fig:comp-winning} and \ref{fig:comp-income} display the distributions of cosine similarity scores between right-wing constituents and left-wing elected politicians, and left-wing constituents and right-wing elected politicians across deciles of winning majority and income. 
We observe that the similarity is lower (close to zero) than in our earlier analysis, when we conducted within constituency comparisons. 
We argue that this is reasonable as elected politicians are more likely to have common concerns with their constituents than with citizens that they do not represent and cannot vote for them (\ie because they belong to a different constituency).



\section{Style-related similarities}
\label{sec:liwc_similarities}
The previous section identified some trends in how elected politicians and their constituents differ in the content they post online depending on the income or the winning majority of the constituency.
We now investigate whether there are also differences that pertain more to the tone or style than to the content.
This is, while they might discuss the same topic, \eg immigration, might differ in the style, choice of words, and tone used. To capture better that nuance, we now use LIWC categories and sentiment analysis.
\begin{table*}[!tb]
    \begin{adjustbox}{max width=\linewidth}
        \begin{tabular}{|c|cccc|cccc|}
            \hline
            & \multicolumn{4}{c|}{\textbf{Winning Majority}} & \multicolumn{4}{c|}{\textbf{Income}} \\
            \hline
            & \multicolumn{2}{c|}{\textbf{USA}} & \multicolumn{2}{c|}{\textbf{UK}} & \multicolumn{2}{c|}{\textbf{USA}} & \multicolumn{2}{c|}{\textbf{UK}} \\
            \hline
            & \multicolumn{1}{c|}{\textbf{Left-wing}} & \multicolumn{1}{c|}{\textbf{Right-wing}} & \multicolumn{1}{c|}{\textbf{Left-wing}} & \textbf{Right-wing} & \multicolumn{1}{c|}{\textbf{Left-wing}} & \multicolumn{1}{c|}{\textbf{Right-wing}} & \multicolumn{1}{c|}{\textbf{Left-wing}} & \textbf{Right-wing} \\
            \hline
            \textbf{Tone} & \multicolumn{1}{c|}{-0.0671} & \multicolumn{1}{c|}{-0.0236} & \multicolumn{1}{c|}{-0.0822} & 0.0941 & \multicolumn{1}{c|}{-0.0278} & \multicolumn{1}{c|}{-0.0674} & \multicolumn{1}{c|}{-0.011} & -0.1129 \\
            \hline
            \textbf{Authentic} & \multicolumn{1}{c|}{-0.02} & \multicolumn{1}{c|}{-0.0802} & \multicolumn{1}{c|}{0.0213} & 0.0297 & \multicolumn{1}{c|}{-0.0299} & \multicolumn{1}{c|}{0.0494} & \multicolumn{1}{c|}{-0.0116} & -0.0405 \\
            \hline
            \textbf{Analytic} & \multicolumn{1}{c|}{-0.0317} & \multicolumn{1}{c|}{-0.0051} & \multicolumn{1}{c|}{-0.0075} & 0.0214 & \multicolumn{1}{c|}{0.0431} & \multicolumn{1}{c|}{0.0804} & \multicolumn{1}{c|}{0.0268} & -0.0131 \\
            \hline
            \textbf{Clout} & \multicolumn{1}{c|}{0.0143} & \multicolumn{1}{c|}{-0.1057} & \multicolumn{1}{c|}{-0.0282} & -0.0272 & \multicolumn{1}{c|}{0.0092} & \multicolumn{1}{c|}{0.0364} & \multicolumn{1}{c|}{0.0505} & -0.0045 \\
            \hline
            \textbf{Linguistic} & \multicolumn{1}{c|}{-0.0093} & \multicolumn{1}{c|}{-0.0009} & \multicolumn{1}{c|}{0.0036} & 0.0209 & \multicolumn{1}{c|}{0.0052} & \multicolumn{1}{c|}{0.0279} & \multicolumn{1}{c|}{-0.0092} & -0.0099 \\
            \hline
            
        \end{tabular}
    \end{adjustbox}
    \caption{Inter-decile range for difference of LIWC categories scores between elected politicians and constituents.}
    \label{tab:liwc-idr}
\end{table*}

\subsection{Psychological similarities}
\label{subsec:liwc_score}
Sylwester et al. found that some LIWC categories\footnote{\ie ``1st person singular pronoun (i)'', ``1st person plural noun (we)'', ``swear words (swear)'', ``positive sentiment (emo$\_$pos), ``negative sentiment (emo$\_$neg)'',``Anxiety (emo$\_$aux)'', ``Feeling (feel)'', ``tentative (tentat)'', ``certainty (certitude)'', ``achievement (achieve)'', religion (relig)'', and ``death (death)''}
identify political orientation (left and right-wing) in Twitter posts~\cite{sylwester2015twitter}. 
We therefore expect that some LIWC categories might reflect the different ways in which constituents and elected politicians express themselves.
%



To observe the variations in style of discourse between elected politician and constituent, we compute differences between LIWC category scores across constituencies. To see whether these differences are significant, we apply two-sample t-test on LIWC category scores from tweets and posts.
We verify that the underlying distributions are independent, a requirement for two sample t-test. 
We calculate the mutual information score \cite{peng2005feature} for every LIWC category in the Twitter and Nextdoor datasets finding values close to 0 (varying between 0.08 and 0.13), where 0 indicates complete independence.

\begin{figure}[!htpb]
    \begin{center}
        \includegraphics[width=\columnwidth]{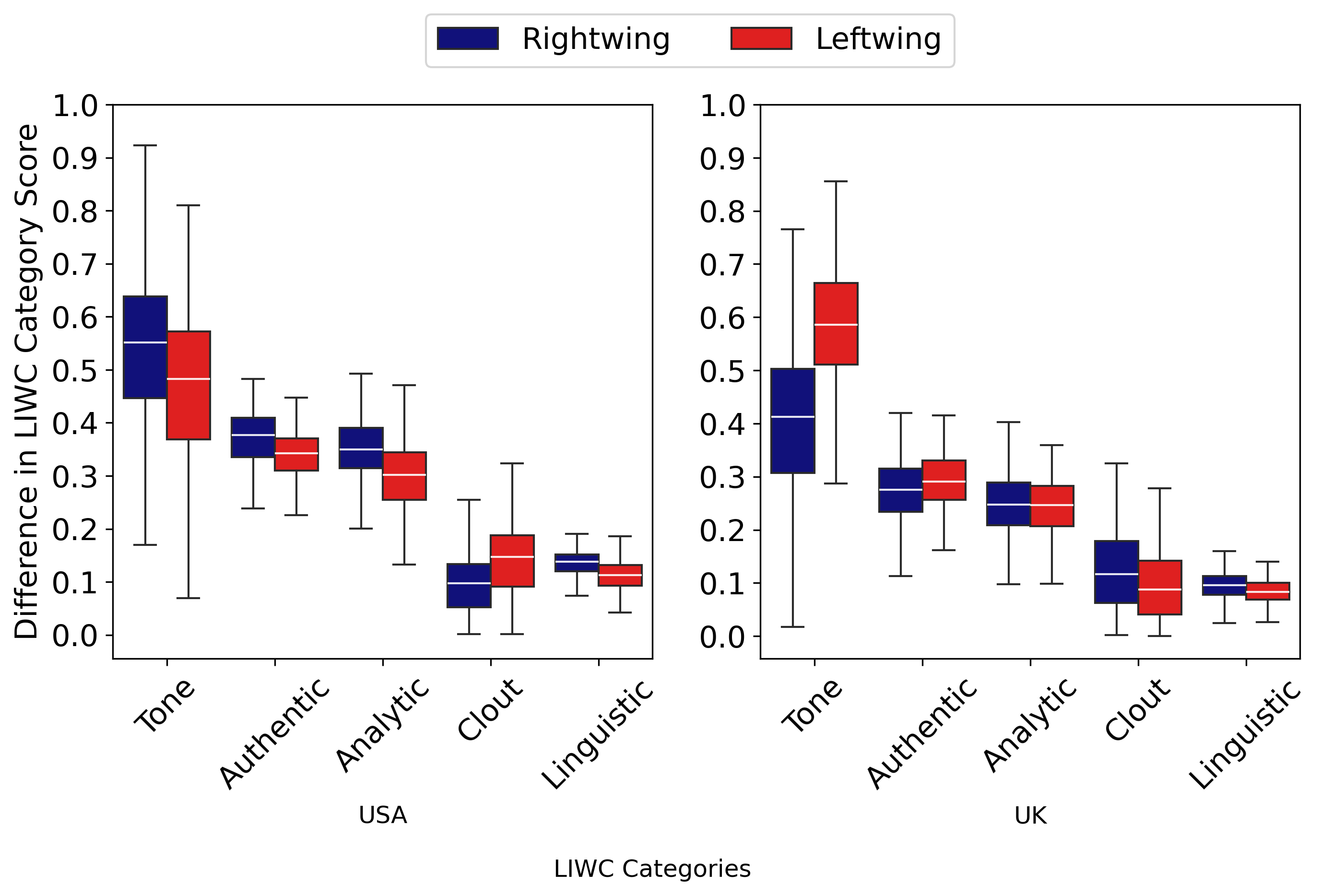}
    \end{center}
    \caption{Distribution of differences of top five LIWC categories score of online discourse between constituents and elected politicians (Scale:0-1)}
    \label{fig:liwc_med_score}
\end{figure}

We find that there are only five LIWC categories (\ie Tone, Analytic, Clout, Authentic, and Linguistic) with statistically significant (\ie $p\geq 0.05$) 
 differences between discourse style of elected politicians and constituents.
These categories cover a large number of subcategories for writing style~\eg \emph{Tone} covers usage of emotions-related words in writings which are further divided into subcategories ~\eg emo\_pos(Positive emotions) and emo\_neg(negative emotions), \emph{Analytic} covers usage of words related formal thinking and reasoning. \emph{Clout} covers words related to leadership and status, \emph{Authentic} covers words related to perceived honesty, and \emph{Linguistic} category refers to the usage of writing structure such as verbs, nouns, and pronouns. 
Figure \ref{fig:liwc_med_score} shows the differences between the five LIWC scores of the constituents and their respective elected politicians of each constituency.

\pb{Elected politicians have a similar discourse style to their constituents regardless of whether the constituency is right or left-wing.}
Constituents and their elected politicians generally have similar styles of online discourse.
%
We find that the differences between right-wing and left-wing constituencies are small even when significant.
We only find a relatively large difference between right and left-wing for the differences in the LIWC category of Tone (median difference of $0.57$ and $0.48$, and $0.42$ and $0.58$ in the USA and UK respectively). 
We find that elected politicians use more negative tone (Tone scores below 0.5 suggest a more negative emotional tone) (median politicians Tone LIWC score of $0.22$ and $0.29$ in USA and UK) than their constituents (median constituents Tone LIWC score of $0.73$ and $0.8$ in USA and UK) in online discourse. We also analyze their tone of online discourse across left-wing and right-wing constituencies and find similar results to aforementioned ones in left-wing and right-wing constituencies for elected politicians ($0.25$ and $0.17$ in USA, $0.21$ and $0.31$ in UK) and constituents ($0.77$ and $0.69$ in USA, $0.78$ and $0.82$ in UK).

We also analyze the categories reported to be a good identifier of political preferences~\cite{sylwester2015twitter}. 
We find that while they might help identify right and left-wing individuals, but scores for these categories are very similar for both constituents and elected politicians.

\pb{Elected politicians of more disputed constituencies have more similar discourse style to constituents.}
Table \ref{tab:liwc-idr} shows the inter-decile range (IDR) of differences of LIWC scores (for the five significant  categories) between elected politicians and their respective constituents.
Depending on size of the winning majority, we observe that the IDR values of most categories (Tone, Analytic, Authentic, and Clout) are negative in left-wing and right-wing constituencies in the USA and left-wing constituencies in UK, but positive in right-wing constituencies in the UK. 
This trend shows that elected politicians tend to use more a more similar style of discourse to their constituents when the margin of victory is low.

A potential hypothesis is that politicians whose victory is not secure are try to be more empathetic with potential voters in order to secure their votes. The findings here are however in opposition to those in Section~\ref{subsec:semantic} showing a nuanced interplay between constituents and politicians.

\pb{Elected politicians in richer constituencies tend to have more similar style to their constituents.}
We also observe that the similarity in style decreases in lower-income constituencies except for right-wing constituencies in the UK.
Table~\ref{tab:liwc-idr} shows that style similarity decreases in  low-income constituencies except for \emph{Tone and Authentic} categories. We also observe that style similarity is increasing in low-income right-wing constituencies in the UK with negative values of IDR. 

\subsection{Sentiment similarity}
To further our analysis on style differences, we know analyze differences in sentiment. 

\begin{figure}[!tbph]
	\begin{center}
		\includegraphics[width=\columnwidth]{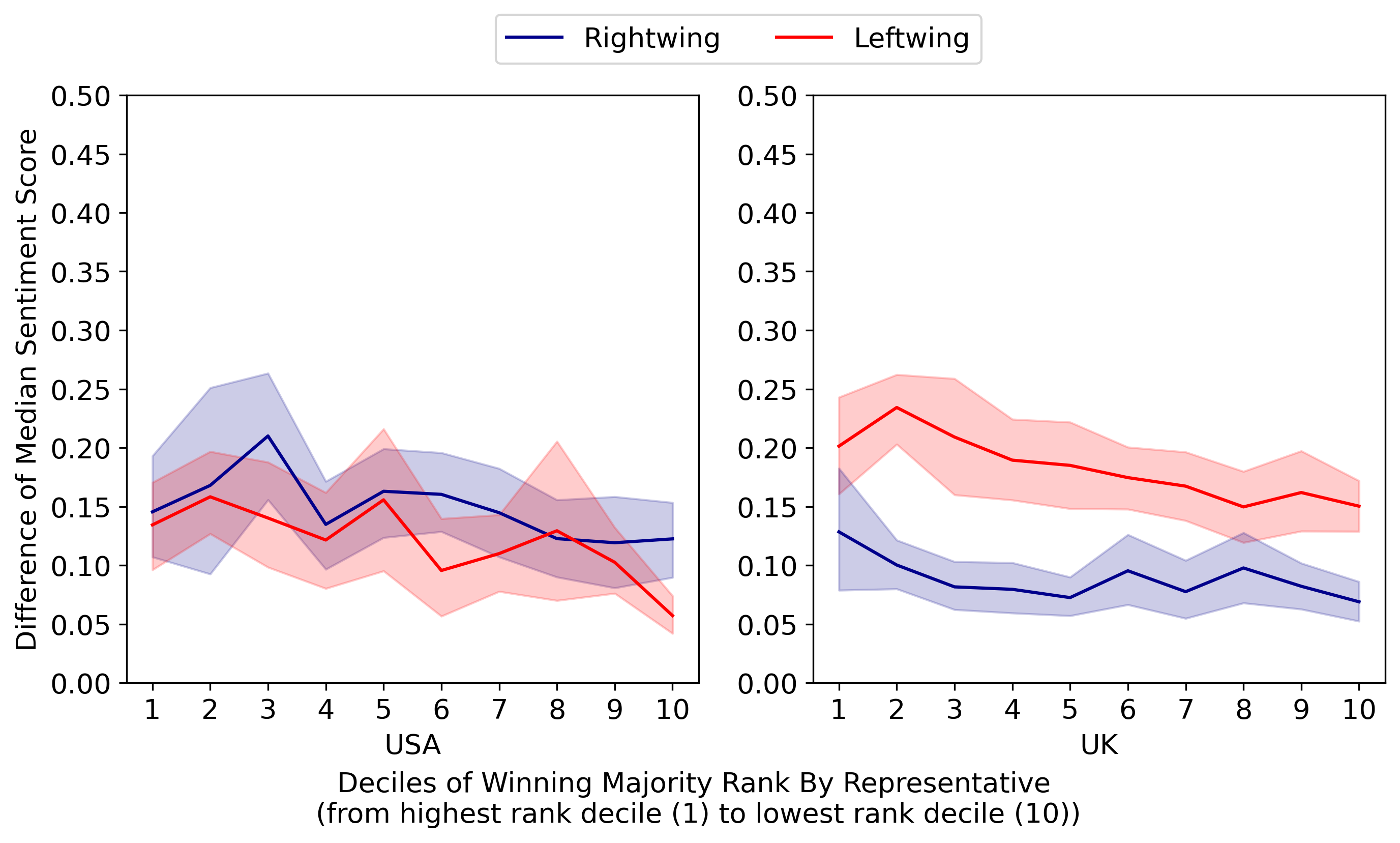}
	\end{center}
	\caption{Distribution of the difference of the median compound sentiment score of elected politicians' tweets on Twitter and constituents' posts on Nextdoor over deciles by the winning voting majority in different constituencies (from higher to lower)}
	\label{fig:sentiment_pol_line_diff}
\end{figure}

\pb{The narrower the majority, the more similar the sentiment between elected politicians and their constituents is.}
Elected politicians of constituencies with narrow majorities have more similar sentiments to their constituents. 
Figure \ref{fig:sentiment_pol_line_diff} shows a decreasing (absolute) sentiment difference  between elected politicians and their respective constituents, as the  winning majority narrows. This is true for both countries and across the political spectrum (see Table~\ref{tab:idr_cos} for IDR values).
This is aligned with our LIWC categories analysis and again opposed to the findings in the previous section (Section~\ref{subsec:semantic}).



%

%

\pb{Elected politicians from poorer constituencies have more similar sentiment to their constituents.}
Elected politicians from poorer constituencies have more similar sentiments to their constituents.

\begin{figure}[!h]
	\begin{center}
		\includegraphics[width=0.95\columnwidth]{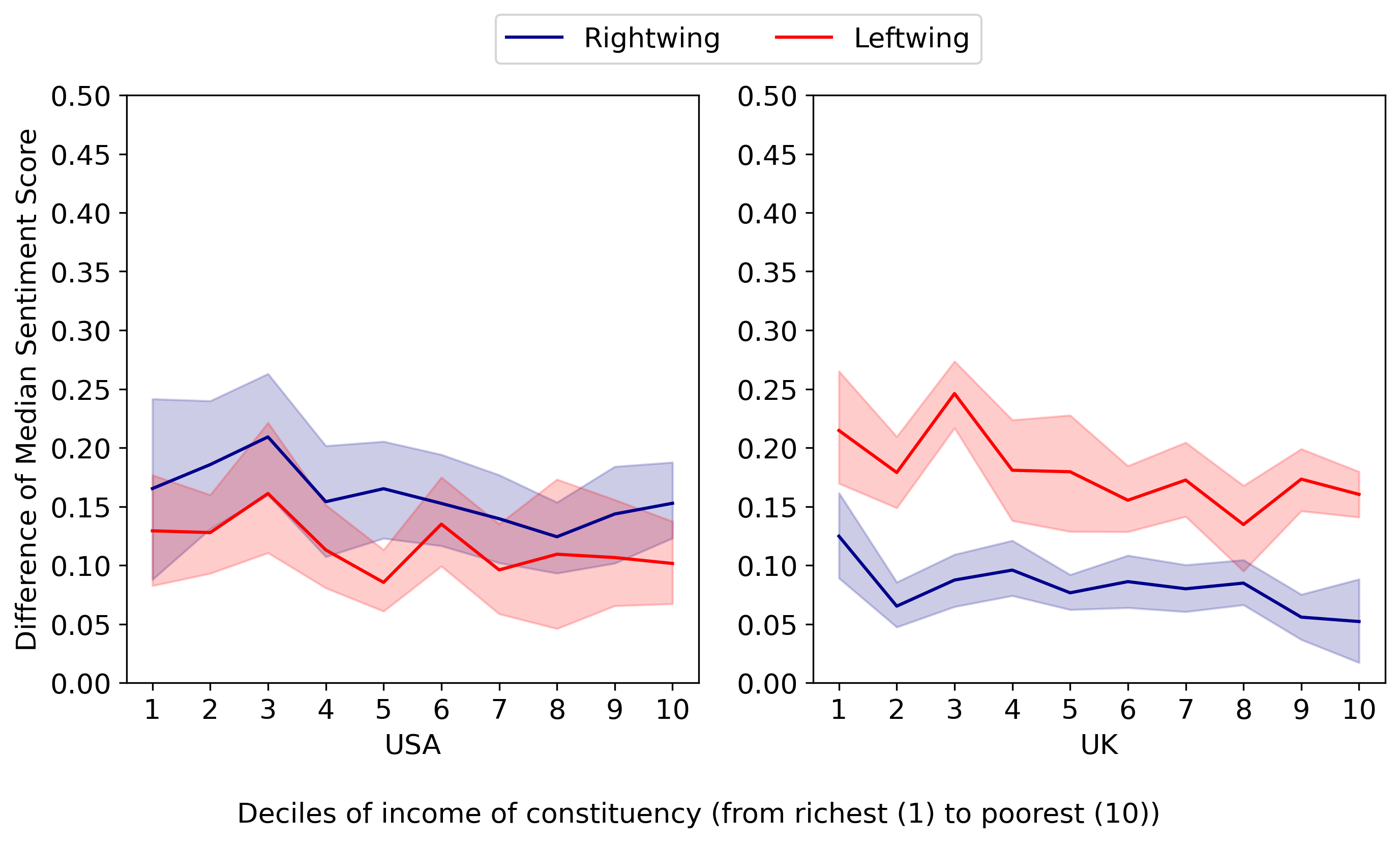}
	\end{center}
	\caption{Distribution of difference of median compound sentiment score of elected politicians' tweets on Twitter and constituents' posts on Nextdoor over deciles by income levels in different constituencies (from richest to poorest).}
	\label{fig:sentiment_income_line_diff}
\end{figure}

Figure \ref{fig:sentiment_income_line_diff} shows how the (absolute) sentiment difference between elected politicians and  their respective constituents decreases with the level of income of the constituency (see  Table ~\ref{tab:idr_cos} for IDR values).
This is in accordance with our earlier findings in Section~\ref{subsec:semantic}: constituents and their elected politicians are more similar in both content and style in poorer neighborhoods. 
%

\section{Limitations}
\label{sec:limitation}
\pb{Comparing Twitter with Nextdoor.} 
There is a particularly important difference between Twitter and Nextdoor:
Twitter welcomes general content and Nextdoor focuses on local issues.
Politicians can run  political campaigns in Twitter,
but political campaigns are not allowed on Nextdoor \cite{nextdoor-pol-adv}.
This might result in relatively low similarities between constituents and elected representatives.
This difference is constant though, it will affect rather absolute than relative values, allowing us 
to compare how similar constituents and their respective elected politicians are across constituencies. 
We also compute similarities between politicians and constituencies that did not elect them. We see that the similarity is consistently and substantially lower.

\pb{Data Representivity and bias.}
\label{data-representivity}
Our data has good geographic coverage and a high correlation with the underlying population.
While we miss some constituencies (2 in USA, 55 in the UK), this is unlikely to have a large effect in the results. 
We do not know, however, how representative of neighborhood the Nextdoor data is, \ie  we have good coverage of regions with different levels of income, but we have no intra-neighborhood visibility.
Our Nextdoor data might suffer from bias towards individuals with certain socioeconomic status (\eg richer individuals might be more likely to use social media~\cite{richer-socmedia-pew}). 
This is a common challenge in quantitative research with social media data 
and studies frequently rely on the level of usage as a proxy of representativity~\cite{chetty2022social, bailey2018economic, bailey2020social, jones2013inferring}. 

\section{Related Work}
\label{sec:relatedwork}

Both voters and  politicians have used social media to discuss politics.
The literature has extensively analyzed the political debate on social media and specially in  Twitter, finding a trend of growing political polarization~\cite{garimella2017long,esteve2022political}.
A number of works have also studied how politicians use social media for political purposes~\cite{parmelee2023personalization, agarwal2019tweeting}.

Our work differs from these in that we focus on the relationship between politicians and voters.
As such, closer to our work is the research on how politicians engage with voters on social media~\cite{mediabaxter2016members,hofmann2013makes,tromble2018great}.
\citeauthor{graham2013between}~\cite{graham2013between} analyzed tweets from 416 UK candidates
and \citeauthor{tromble2018great}~\cite{tromble2018great} analyzed tweets from 992 elected politicians (418 American, 434 British, and 140 Dutch).

Our work again differ in that instead of looking at the engagement of politicians with a subset of individuals who might belong to a different constituency, we explicitly pair politicians and constituents across USA and UK. To do this we collect a comprehensive dataset from Twitter, for elected politicians, and Nextdoor, for their constituents.



To the best of our knowledge, there has been only one prior quantitative study of Nextdoor.
Iqbal et al~\cite{iqbal2023lady} show that Nextdoor can be used to predict socioeconomic parameters of their neighborhoods. Nextdor maps users to geographical neighborhoods, and similarly to this work, we leverage the geographical tagging of Nextdoor users and group them into political constituencies. This allows us to ensure that we compare elected politicians with residents of their constituency.

\section{Conclusion}
\label{sec:conclusion}
This paper examined how similar elected politicians and their constituents are.
We examined this in terms of the similarities in content and style of their online discourse.
To do this, we conducted a large-scale analysis where we collected 21.4 Million posts from 433 USA and 595 UK constituencies from Nextdoor, and 5.6 Million tweets from the accounts of the elected politicians of the same constituencies. 
We found that elected politicians tend to be equally similar to their constituents in terms of discourse content and style regardless of whether a constituency elects a right or left-wing politician. 
The size of the electoral victory and the level of income of a constituency showed a nuanced picture. We found that narrower electoral victories were associated with a more similar discourse style (both in terms of LIWC categories and sentiment). We found the opposite for the discourse's content: the larger the victory the more similar the content tended to be.
Poorer constituencies showed more similarity in terms of content. In terms of style, poorer constituencies showed a more similar sentiment and the opposite was true for the psychological text traits (\ie measured with LIWC categories).

\section{Ethical Statement}
\label{sec:ethics}
This research study has been approved by the Institutional Review Board (IRB) at the researchers’ institution.
The authors have no competing interests or funding that could undermine this research.
We employ users' public post records from Nextdoor to study their conversations. 
Nextdoor data is public, as there is the expectation that strangers can view the posts~\cite{townsend2016social}.
Upon collection, we anonymize the data before use and store it in a secure silo.
We aggregate our data and analyze it at a constituency level to prevent user identification. We discard any user-level information. 
Our work does not share or redistribute Nextdoor content, as per Nextdoor's Terms of Service. Importantly, web crawling is legal for non-commercial research in the UK \cite{scraping-uk} and the USA \cite{scraping-usa}, where the data collection is performed. 

We also gather tweets from elected politicians in the USA and UK using Twitter Academic API access. This data is considered public data because it is accessible publicly and anyone can interact with this data even without permission of the original author of the tweet. Townsend et al. have discussed in this situation in great detail in case 5~\cite{townsend2016social}.
 \bibliography{aaai23}   
\section{Appendix}


\end{document}